%% file: chiral.txt
\par
\input phyzzx\input mydef
\def\pr{\prime}
\def\e{\adveq\eqno{\rm (\chapterlabel.\the\equanumber)}}
\date{March, 2009}
\date{March, 2009}
\titlepage
\title{Chiral Algebras for Superconformal Interacting Bosons}
\author{Doron Gepner}
\line{\hfill Department of Particle Physics\hfill}
\line{\hfill Weizmann Institute\hfill}
\line{\hfill Rehovot 76100, Israel\hfill}
\abstract
A sort of calculus is developed to find the chiral algebras of
$N=2$ superconformal interacting bosonic models. Many examples
are discussed. It is shown that the algebras share a common structure,
which we call almost Landau Ginzburg. For one or two generators,
the number of relations is equal to the number of generators and they
are algebraically independent.
\endpage
\mysec{Introduction}
String theory has attracted considerable attention due to its
potential to unify all the forces in nature, including gravity, along
with giving the phenomenology of the known particle spectrum.
In ref. \REF\DG{D. Gepner, Nucl. Phys. B296 (1988) 757}\r\DG,
heterotic--like, spacetime supersymmetric string theories were 
described starting from any $N=2$ superconformal models in two
dimensions, with the central charge $c=3 r$, where $2r$ is the 
number of compactified dimensions. For four dimensional 
compactifications there is a gauge group which includes $E_6$, with
chiral fermions in the $27$ and $\bar{27}$ representations, and is
thus capable of realistic physics.  
The constructions discussed were
based on the tensor products of $N=2$ superconformal minimal models.
So, for example, the theory $1^1 16^3_E$ (one $k=1$ and three $k=16$
minimal models, with the exceptional modular invariant) predicts
three generations of particles in the $27$ of $E_6$, 
\REF\Three{D. Gepner, Princeton preprint (1988), hep-th/9301089}
\r\Three. 

A geometric description of these theories was found in ref. \REF\CY{D.
Gepner, Phys. Lett. B199 (1987) 380}\r\CY, where it was shown that the
spectrum of some such theories is the one expected from compactification
on complex manifolds with vanishing first Chern class, so called 
Calabi--Yau manifolds, originally discussed by Candelas et al.
\REF\Cand{P. Candelas, G.T. Horowitz, A. Strominger and E. Witten,
Nucl. Phys. B258 (1985) 46.}\r\Cand. So, for example, the theory 
$3^5$ has the spectrum expected from the quintic hypersurface in 
$CP^4$.

Much of the geometry is encoded in the so called chiral algebra
of the $N=2$ superconformal models. These are fields that obey
$\Delta=Q/2$, where $\Delta$ is the dimension and $Q$ is the $U(1)$
charge. (similarly, for the right movers $\bar\Delta=\bar Q/2$.)
These fields can be multiplyed by one another since the OPE is
non--singular. I.e., if $C_i$ and $C_j$ are two such fields, we define
the product chiral field, $C_k=C_i C_j$ as
$$C_k(w)=\lim_{z\rarrow w} C_i(z) C_j(w).\e$$
Since the OPE is associative, this defines an algebra.

A large class of $N=2$ superconformal models was put forwards 
by Kazama and Suzuki,
based on interacting bosonic models (IBM) \REF\KS{Y. Kazama and H.
Suzuki, Mod. Phys. Lett. A4 (1989) 235.}\r\KS, and their Hilbert
space was described in ref. \REF\CG{R. Cohen and D. Gepner, 
Mod. Phys. Lett. A vol. 6, No. 24 (1991) 2249.}\r\CG. Our aim
in this paper is to describe the chiral algebra of such models.

We develop a sort of ``calculus" to write down the chiral algebra
of the models, and we discuss it explicitly in many examples.
We hope that these results will serve to clarify the geometrical 
nature of the models.

We find a common structure for the chiral algebras of the models.
This we call almost Landau Ginzburg (ALG). It is easiest to describe
when there are one or two generators, there the number of relations
in the algebra is equal to the number of generators and the relations
are algebraically independent. With more generators the structure
is more complicated, for example for three generators there are four
relations, but it is also common to all the models discussed.  
\par
\mysec{$N=2$ Superconformal interacting bosons}

The $N=2$ interacting bosonic models of Kazama and Suzuki are 
defined as follows \r\KS. We take a set of vectors of length three,
which we denote by $\gamma_i$. We assume that the scalar products 
of these vectors are either $0$ or $1$,
$$\gamma_i \gamma_j=\Gamma_{ij}=0{\ \rm or\ }1,\e$$
for $i\neq j$ and $\gamma_i^2=3$.

We define the super stress tensors $G_{\pm}(z)$ by
$$\eqalign{G_+&=\sum_\gamma f_\gamma :e^{i\gamma\phi}:,\cr
         G_-&=\sum_\gamma f_\gamma^\dagger :e^{-i\gamma\phi}:,\cr}\e$$
where $\phi$ is a set of canonical free bosons with the same dimension
as $\gamma$, and $f_\gamma$ are some complex numbers. The 
$f_\gamma$ are also multiplied by a cocycle operator, which since it
does not enter into our discussion here, for the most part, we will
ignore.
The rest of the elements of the $N=2$ superconformal algebra are then
dictated by the OPE of the algebra, and they are,
$$\manyeq{J&={i\over2}\sum_\gamma |f_\gamma|^2 \gamma\partial\phi,\cr
  T&=-{1\over4}\sum_\gamma |f_\gamma|^2 (\gamma\partial\phi)^2+
     \sum a(\alpha) :e^{i\alpha\phi}:,\cr}$$
where
$$a(\alpha)={1\over2} \sum_{\gamma,\gamma^\pr,\gamma\gamma^\pr=1\atop
            \gamma-\gamma^\pr=\alpha} f_\gamma f_{\gamma^\pr}^\dagger.
\e$$
Importantly, for every such matrix $\Gamma$, the $N=2$ algebra closes,
and we obtain an $N=2$ superconformal field theory, provided that
the following relations hold,
$$\sum_j \Gamma_{ij} x_j=2,\qquad{\rm where\ } x_j=
|f_{\gamma_j}|^2,\e$$
and the central charge is
$$c={3\over2} \sum_i x_i.\e$$

Other models exist where we let the off diagonal elements of $\Gamma$
be $0$ or $\pm1$. However, these do not always lead to a consistent
$N=2$ superconformal algebra, and will not be discussed here.
We can denote each such matrix $\Gamma$, which defines the model, by an
unoriented graph of $n$ points, $n=\dim \gamma$, where we connect
the two points $i$ and $j$  if $\Gamma_{ij}=1$, and leave it 
unconnected if $\Gamma_{ij}=0$. Thus, each graph uniquely defines 
an interacting bosonic model (IBM). Clearly, isomorphic graphs lead
to the same theory, and if the graph is disconnected, the theory is
a tensor product of each of the theories associated with each connected
piece. Thus, it is enough to consider only connected unoriented graphs.

As a n example consider the complete graph, 
$\Gamma_{ij}=2\delta_{ij}+1$, for $i,j=1,2,\ldots,n$. 
It is easy to see that $x_i=2/(n+2)$ and thus form eq. (2.6),
the central charge is given by
$$c={3 n\over n+2},\e$$
which is the central charge of the $n$'th minimal model. We 
conclude that the complete graphs corresponds to the minimal models.

The free boson operator, that we take, must be local with respect to 
the $N=2$ generators, $T$, $G_{\pm}$ and $J$, in the Neveu--Schwarz
sector (NS). Any operator in the
free boson theory is of the form
$a=e^{ip\phi}$ (times derivatives of $\phi$), that is the momenta is $p$.
Then, locality with $G_+$ is obeyed if the momenta $p$ lies on the
dual lattice of $\Gamma$,
$$p\gamma_i={\rm integer},\e$$
and in the Ramond sector (R), this condition becomes, $p\gamma_i=
{\rm integer}+\half$. Denoting the dual lattice generators by 
$\zeta_i$, 
$$\gamma_i\zeta_j=\delta_{ij},\e$$
we conclude that each momenta can be written by 
$$p=\sum_{i} \epsilon_i \zeta_i,\e$$
where $\epsilon_i$ are some integers, in the NS sector.
It is enough to consider only the NS sector since the R sector is 
isomorphic to it ref. 
\REF\SS{N. Seiberg and A. Schwimmer, Phys. Lett. B184 (1987) 191.}
\r\SS. We conclude that the free boson theory is defined on the
dual lattice to $\Gamma$, denoted by $\Gamma^*$. The operator $T_A^0$ will mix
fields with equal dimension in the free theory. Thus every field
is of the form
$$b=\sum_{r} :e^{ip_r\phi} e^{i p_r\bar\phi}: O_r,\e$$
where $O_r$ are a product of some derivatives of $\phi$ and 
$\bar \phi$, all
of the same dimension in the free boson theory. 
We assumed here that the bosonic theory is left right symmetric,
i.e., the left momenta and the right momenta are equal. In the
sequel, we will omit the right moving part, as it is always the same
as the left moving part.
In particular
$p_r^2/2$ are all the same modulo an integer.

The primary fields in an $N=2$ algebra, in the NS sector,
are the fields $\theta$
that obey the relation,
$$G_{\pm}(z) \theta(w)=O[(z-w)^{-1}],\e$$
and the chiral fields are the primary fields which obey also,
$$G^+(z) \theta(w)=O(1).\e$$
It follows that the fields $\theta=:e^{ip\phi}:$ where,
the momenta is $p=\sum_r \epsilon_r \zeta_r$, and $\epsilon_r=0,\pm1$,
will be primary fields.  
Clearly, when $A$ is not a minimal model,
there are infinite number of primary fields, not all of this form.
However, these are guaranteed to be primary fields.
The chiral fields will be given by the same expression, where 
$\epsilon_i$ are limited to be zero or one, $\epsilon_i=0,1$.
From the left--right symmetry these are
fields which are both left and right chiral. There are no fields
which are left chiral and right anti--chiral.
Clearly, there might be other chiral fields, not of this form.
Some examples will be discussed in the sequel. We call the set
of this fields as the basic chiral fields, which are guaranteed
to be chiral for every such $\epsilon_i$. We will denote such a field
by its $\epsilon_i$, $\theta=(\epsilon_1,\epsilon_2,\ldots,\epsilon_n)$.
From eq. (2.3), the $U(1)$ charge of the field $\theta$ is
$$q=\half\sum_r \epsilon_r x_r.\e$$
The maximal chiral field is given by the field 
$C_{\rm max}=(1,1,\ldots,1)$, and it will have the $U(1)$ charge 
$c/3=\half\sum x_i$.

Let us turn now to the Hilbert space of the theories, following ref.
\r\CG. Denote by $T_0$ the stress tensor
of the free theory of $n$ bosons,
$$T_0= -{1\over2} \sum_{i=1}^n (\partial\phi_i)^2.\e$$
The key to solving the theory is the observation that $T_B=T_0-T$
and $T$ commute, i.e., their OPE is nonsingular. 
We denote the IBM with the stress tensor $T$ as the $A$ theory.
Thus, the theory $A$,
tensored with the $B$ theory, gives a free boson theory. This implies,
in particular, that the operators $T^0_A$ and $T^0_B$, the dimensions,
commute, and thus can be diagonalized simultaneously. It follows, that
each operator in the free theory, along with its mixings by $T_A$
can be diagonalized to give a field in the $A$ theory, since
$T_A$ is the identity in the $B$ theory, and the problem
of finding the spectrum, reduces to a finite dimensional 
diagonalization of the action of $T_A^0$. 

For the basic primary fields which are of the type 
$\theta=(\epsilon_1,\epsilon_2,\ldots,\epsilon_n)$ such
that $|\epsilon_i|\leq1$, the mixing matrix, under $T_0^A$ is particularly
simple to describe. The field $\theta$ will mix with fields of the
form $\theta_{ij}=\theta+\gamma_i -\gamma_j$, where $\gamma_i\gamma_j=1$ and
$\epsilon_i-\epsilon_j=2$. From the expression for $T_A$, eq. (2.3),
the diagonal elements of the mixing matrix are,
$$M(\theta,\theta)={1\over4} \sum_r x_r \epsilon_r^2,\e$$
and the off diagonal elements are    
$$M(\theta,\theta_{ij})={1\over2}\sqrt{x_i x_j}.\e$$
Diagonalizing the matrix $M$ we get the dimensions of the field 
$\theta$, while the eigenvectors denote the specific combinations 
which are fields in the $A$ theory.
Other fields will involve derivatives of $\phi$ in their mixings,
and thus the matrix describing $L_0^A$ will be much more complicated.
See ref. \REF\CG{J.P. Conlon and D. Gepner, 
Phys. Lett. B548 (2002) 102.}\r\CG\ for explicit calculations
of such mixings. Here, however, we shall not perform such mixings 
explicitly.

Let us assume that we have a graph of the type,
$$\Gamma=\pmatrix{3&1&0&0&\ldots&0&0&\ldots\cr
                 1&3&1&1&\ldots&1&0&\ldots\cr
           \ldots&\ldots&\ldots&\ldots&\ldots&\ldots&\ldots&\ldots\cr 
           \ldots&\ldots&\ldots&\ldots&\ldots&\ldots&\ldots&\ldots\cr
                 },\e$$
where in the second row we have $r+1$ ones, 
for some $r$.
From eq. (2.5) to get the $x_i$ we need to solve the equation,
$\Gamma_{ij} x_j=2$. Denote by $y$,
$$y=\sum_{i=3}^{r+2} x_i.\e$$
Then, eq. (2.5) implies that,
$$\manyeq{3x_1+x_2&=2,\cr
          3x_2+x_1+y&=2.\cr}$$
Solving it we find that 
$$\manyeq{x_1&={4+y\over 8},\cr
  x_2&={4-3 y\over8}.\cr}$$
Now, consider the field,
$$a_1=(1,-1,0,0,\ldots,0).\e$$
It mixes only with the field 
$$a_2=a_1+\gamma_2-\gamma_1=(-1,1,1,\ldots,1,0,0,\ldots,0),\e$$
where there are $r+1$ ones in $a_2$. 
To get the dimensions of this field we need to find
the eigenvalues of the mixing matrix, eq. (2.16,2.17). A short
calculation shows that this two by two matrix is given here by,
$$\manyeq{M_{11}&={1\over4}\sum_i x_i a_{1i}^2={1\over4}(1-y/4)\cr,
  M_{22}&={1\over4} \sum_i x_i a_{2i}^2={1\over4} (1+3 y/4),\cr
  M_{12}&=M_{21}={1\over2} \sqrt{x_1 x_2}={1\over16} 
\sqrt{(4-3 y)(4+y)}.\cr}$$
The matrix we need to diagonalize, $M$, is given by
$$M=\pmatrix{M_{11}&M_{12}\cr M_{21}&M_{22}\cr}.\e$$
A quick calculation shows that the eigenvalues of this matrix are
$\Delta=\half$ and $y/8$,
with the eigenvectors
$$\{ {\sqrt{16-8 y-3 y^2}\over y+4},1\},\e$$
and 
$$\{ {\sqrt{16-8 y-3 y^2}\over 3 y-4},1\},\e $$
Thus the field $\phi=c_1 a_1+c_2 a_2$ has the dimension $\Delta=y/8$,
with the $c$'s given by eq. (2.26). 
Since the $U(1)$ charge of the field $a_1$ is
$\half(x_1-x_2)=y/4$ it follows that the field $\phi$ is a chiral
field, since it obeys $Q=2\Delta$.
We conclude that the field $a_1$, along with its mixings, 
for any matrix of the type $\Gamma$
given by eq. (2.18) is chiral, despite the fact that it is not given
by the canonical type $\epsilon_i=0,1$.
The square of this field is given by
$$a_1 a_2=(0,0,1,1,\ldots,1,0,\ldots,0),\e$$
where there are $r$ ones in this field.
To summarize, each time that the graph of $\Gamma$ has a ``tail",
there will be the chiral field associated to this tail given by 
$a_1$, with the $U(1)$ charge $Q=y/4$.

\mysec{The chiral algebras}
\par
The problem of calculating the chiral fields of a given IBM model
is quite hard, since we do not know up to what dimensions we need to 
go in order to get all the fields. Here, we will make
a simplifying ansatz that all the chiral fields are of the form
as follows. Either, they are given by the basic chiral fields
of the form $(\epsilon_1,\epsilon_2,\ldots,\epsilon_n)$, such that
$\epsilon_r=0$ or $1$, or, for each tail of the graph, 
as explained above, we have the field $(1,-1,0,\ldots,0)$ where 
positions $1$ and $2$ refer to the tail.
As we shall see in the examples in the sequel this ansatz leads to
consistent results. There are examples where this ansatz is not 
justified, and there are additional chiral fields, but these will not
be discussed here.

We shall divide the relationships in the algebra to two kinds. 1) Field
identifications. These we will get using the field identifications in
the chiral algebra. 2) Zero relations. These we will get from the missing 
chiral fields with the maximum $U(1)$ charges. To get
the field identifications we develop a kind of calculus, where we combine
the field identifications with the associative products in the 
chiral algebra. The zero relations we then get from the missing fields at the 
top of the algebra.

Before proceeding it will be useful to work out a simple example.
Consider the graph which is the Dynkin diagram of the algebra $A_n$.
Accordingly, we will call this theory $A_n$.
The matrix of scalar products is given by the $n\times n$ matrix,
$$\Gamma_n=\pmatrix{3&1&0&0&\ldots&0&0&0\cr
                 1&3&1&0&\ldots&0&0&0\cr
            \ldots&\ldots&\ldots&\ldots&\ldots&\ldots&\ldots&\ldots\cr
                 0&0&0&0&\ldots&1&3&1\cr
                 0&0&0&0&\ldots&0&1&3\cr}.\e$$
The graph $\Gamma_n$ is $Z_2$ symmetric under the exchange of the 
point $r$ with the point $n+1-r$. Define the field 
$\theta_r=(0,0,\ldots,1,0,\ldots,0)$, where the $1$ is in the $r$ 
position. We will, thus, assume that the fields $\theta_r$ and 
$\theta_{n+1-r}$ are identical in the IBM. That is, they differ by a 
field identification. It should be possible to justify this assumption
by showing that the field $(0,0,\ldots,1,0,\ldots,0,-1,0,\ldots,0)$,
times derivatives of $\phi$,
where the $1$ and $-1$ are in the $r$th and $n+1-r$th place 
respectively, has zero dimension under $T^0_A$. Since this calculation
involves derivatives of $\phi$, we shall not perform it here.

Let us consider first the simple example of $A_3$. The solution
to eq. (2.5) is
$$x=(4,2,4)/7.\e$$
The central charge is, eq. (2.6),
$$c={3\over2}\sum x_i=15/7,\e$$
which is the central charge of the $5$th minimal model. Thus, we
expect to find the chiral algebra of the $k=5$ minimal model.
This will provide a check for the consistency of our method.
We have two generators for the chiral algebra,
$$x_1=(1,-1,0)=(-1,1,1),\qquad{\rm and\ } z_1=(0,1,0).\e$$
Thus $x_1^2=(1,-1,0)\times (-1,1,1)=(0,0,1)=(1,0,0)$ 
(we simply add the momentas in the product) 
and we find the relation
$$x_1 z_1=(1,0,0)=x_1^2,\e$$
from which we conclude that $z_1=x_1$ (we assume for the field 
identifications that if $ab=ac$ then $b=c$). We thus have one
generator $x_1$. For the zero relation, we note that the maximal 
chiral field is $x_1^5$ of $U(1)$ charge $c/3$. Thus, we are missing
the field $x_1^6$. We conclude that $x_1^6=0$ is the only zero relation,
and the algebra is
$$C\approx {C[x_1]\over (x_1^6)},\e$$
where $(a)$ denotes the ideal generated by $a$, and $C[x_1]$ denotes
all the polynomials of $x_1$ with coefficients which are 
complex numbers. This is exactly the correct chiral algebra of the 
minimal model, confirming our method.

In general, it is convenient to represent the chiral algebra by its Poincare
polynomial, defined by 
$$P(t)=\sum_{\cal C} q^{J_0},\e$$
where $\cal C$ is the chiral algebra generators and $J_0$ is the $U(1)$ 
charge. Due to the transpose property, the Poincare
polynomial obeys,
$$P(t)=t^{c/3} P(1/t).\e$$
The Poincare polynomial is a convenient way to describe the algebra.
For our minimal model example, $A_3$, the polynomial is
$$P(t^7)={1-t^6\over 1-t},\e$$
as is well known. In general, if the degree of the generators (their
$U(1)$ charge times $N$) is $a_1,a_2,\ldots,a_r$ and the degrees
of the relations is $b_1,b_2,\ldots, b_r$, and the relations are
algebraically independent, then the Poincare polynomial is
$$P(t^N)=\prod_{j=1}^r {1-t^{b_j}\over 1-t^{a_j}}.\e$$
As we shall see, for one and two generators the Poincare polynomials
of the IBM under discussion will always be of this form.

The example of $A_3$ can be generalized to any complete graph
of $n$ points which is missing one line. Here, the scalar product
matrix is,
$$\Gamma_{i,j}=2\delta_{i,j}+1-\delta_{i,1}\delta_{j,n}-
\delta_{i,n}\delta_{j,1},\e$$
and $i,j=1,2,\ldots,n$. The missing line is between $1$ and $n$.
The case of $n=3$ is $A_3$ discussed above. We need to solve the
equation $\Gamma_{ij} x_j=2$, eq. (2.5). Denoting $x_1=x_n=x$ and
$x_2=x_3=\ldots x_{n-1}=y$ we find the solution,
$$y={2 \over n+4},\qquad x={4\over n+4}.\e$$
Thus, the central charge is given by
$$c={3\over2}\sum_i x_i={3 N\over N+2},\e$$
where $N=n+2$. This is exactly the central charge of the $N$th 
minimal model. We conclude that the complete graph less a line is 
the $(n+2)$th minimal model where $n$ is the number of points.

We want to see if, by our method, we get the correct chiral algebra.
Denote by $a_2=(1,0,0,\ldots,0)=(0,0,\ldots,0,1)$ the first generator
of degree $2$, and by $b_1=(0,1,0,\ldots,0)=(0,0,1,0,\ldots,0)=
\ldots=(0,0,\ldots,1,0)$
the second generator associated to the $y$ node. The subscript denotes
the degree of the generators. 
Now, consider the field $\theta=(2,1,0,\ldots,0)$. It is identical to
the field 
$$\theta_1=\theta-\gamma_1+\gamma_2=(0,3,0,\ldots,0,1).\e$$
We can write $\theta$ with the generators as
$$\theta=a_2^2 b_1,\e$$
and similarly,
$$\theta_1=b_1^3 a_2.\e$$
We conclude that we have the relation in the algebra,
$$a_2^2 b_1=a_2 b_1^3,\e$$
from which it follows that
$$a_2=b_1^2.\e$$
Thus, we have one generator only, $b_1$, of degree one. The null generator
relation implies that
$$b_1^{n+3}=0,\e$$
and we get the algebra,
$${\cal C}\approx {C[b_1]\over (b_1^{n+3})},\e$$
which is precisely the correct chiral algebra of the $N$th minimal model.
We conclude that our ``calculus'' works correctly for this example.

Let us now do the example of $A_5$. Here the weights are given by
$x=(5,3,4,3,5)/9$. So the central charge is $c=30/9$, so this is
not a minimal model. The generators in the chiral algebra are,
$$x_2=(1,-1,0,0,0),\qquad {\rm and\ } x_3=(0,1,0,0,0),\e$$
whose $U(1)$ charges are $2/18$ and $3/18$ respectively.
By field identification, $x_2=(-1,1,1,0,0)$ and $x_2^2=(0,0,1,0,0)$.
To get the relation between the generators, we use the field
$$\theta=(2,1,0,0,0)=(2,1,0,0,0)-\gamma_2+\gamma_3=(1,-1,2,1,0).\e$$
The field theta can be written on the l.h.s. as $x_2^2 x_3^3$ and
on the r.h.s. as $x_2^5 x_3$. We conclude that
$$x_2^3=x_3^2.\e$$
Our algebra contains $x_2^j$, for all $j$, along with product by
$1,x_3$. Thus we have all degrees except the degree one,
$18 Q={0,2,3,4,\ldots}$. We conclude that the missing field is at
degree $19$, $18 Q=19$ and so
$$x_2^8 x_3=0,\e$$
is the zero relation. These are the only relations. So, our algebra is
$${\cal C}\approx {C[x_2,x_3]\over (x_2^3-x_3^2,x_2^8 x_3)}.\e$$
This is an example of a algebra with two generators. The Poincare
polynomial is 
$$P(t^{18})={(1-t^6)(1-t^{19})\over (1-t^2)(1-t^3)},\e$$
and we see that the relations in the algebra are algebraically 
independent.

Let us consider now the theory $A_6$. Here the weights are
$$x=(16,10,12,12,10,16)/29$$
and the central charge is
$c/3=38/29$. Again, there are two generators in the algebra,
$$x_3=(1,-1,0,0,0,0),\qquad{\rm and\ } x_5=(0,1,0,0,0,0).\e$$
So, $x_3=(-1,1,1,0,0,0)$ and $x_3^2=(0,0,1,0,0,0)=(0,0,0,1,0,0)$.
The relation in the algebra is, again, obtained from,
$$\theta=(2,1,0,0,0,0)=\theta-\gamma_2+\gamma_3=(1,-1,2,1,0,0),\e$$
which implies that $x_3^7=x_3^2 x_5^3$ from which we conclude
$$x_3^5=x_5^3,\e$$
which is the only field identification. The missing field is  
at degree $29 Q=31$ and so
$$x_5^2 x_3^7=0,\e$$
is the zero relation. It can be seen that these are the only relations.
Thus, the chiral algebra is here
$${\cal C}\approx {C[x_3,x_5]\over (x_3^5-x_5^3,x_5^2 x_3^7)}.\e$$
The relations are algebraically independent and so the Poincare
polynomial is,
$$P(t^{29})={(1-t^{15})(1-t^{31})\over (1-t^3)(1-t^5)}.\e$$

We see from the examples of $A_5$ and $A_6$ that when there are 
two generators, there are exactly two algebraically independent
relations. The Poincare polynomial is then always
$$P(t^N)={(1-t^p)(1-t^q)\over (1-t^a)(1-t^b)},\e$$
where $a$ and $b$ are the degrees of the generators and $p$ and $q$
are the degrees of the relations. We call this structure almost
Landau Ginzburg (ALG). For one generator, of course, we also have this
structure. We will see that all IBM's have the ALG structure, though
with three or more generators the algebra can be more complicated.

Consider, now, the graph which is the Dynkin diagram of the algebra
$D_n$. The $n$ by $n$ matrix of scalar products is
$$\Gamma=\pmatrix{3&1&0&0&\ldots&0&0&0&0\cr
                  1&3&1&0&\ldots&0&0&0&0\cr
                  0&1&3&1&\ldots&0&0&0&0\cr 
       \ldots&\ldots&\ldots&\ldots&\ldots&\ldots&\ldots&\ldots&\ldots\cr
                  0&0&0&0&\ldots&0&1&0&3\cr
                  0&0&0&0&\ldots&0&1&3&0\cr}.\e$$
For $D_5$ the weights are 
$$(24,22,4,30,30)/47\e$$
and the central
charge is $c/3=55/47$. We have now three generators,
$$x_1=(1,-1,0,0,0),\qquad x_{11}=(0,1,0,0,0)\qquad{\rm and\ }
x_{15}=(0,0,0,0,1).\e$$
For the first relation we take the field,
$$\theta=(0,2,1,0,0)=\theta-\gamma_3+\gamma_4=(0,1,-1,2,-1).\e$$
Expanding both sides in terms of the generating chiral fields, we
find $x_1^2 x_{11}^2=x_{11} x_1^{-2} x_{15}$ and so
$$x_{15}=x_1^4 x_{11},\e$$
so we are left with two generators $x_1$ and $x_{11}$. Next, use
the relation,
$$\theta_1=(2,1,0,0,0)=\theta_1-\gamma_2+\gamma_3=(1,-1,2,1,1).\e$$
The l.h.s. is $x_1^2 x_{11}^3$ and the r.h.s. is $x_1^5 x_{15}^2$.
We thus find,
$$x_{11}=x_1^{11},\e$$
and we are left with only one generator, $x_1$. The missing field
is at degree $56$ and thus the chiral algebra is
$${\cal C}\approx {C[x_1]\over (x_1^{56})},\e$$
and the Poincare polynomial is
$$P(t^{47})={1-t^{56}\over 1-t}.\e$$
Note that although there is only one generator, this is not a minimal
model. 

Let us do the example of $D_6$. The weights here are 
$$x_i=(70,36,68,6,80,80)/123.\e$$ 
The central charge is $c/3=170/123$.
We have four generators to begin with,
$$\manyeq{x_{17}&=(1,-1,0,0,0,0)\cr
          x_{18}&=(0,1,0,0,0,0)\cr
          x_3&=(0,0,0,1,0,0)\cr
          x_{40}&=(0,0,0,0,0,1)=(0,0,0,0,1,0)\cr}$$
Our first relation comes from the field,
$$\theta=(0,0,2,1,0,0)=\theta-\gamma_4+\gamma_5=(0,0,1,-1,2,-1).\e$$
Expanding both sides in terms of the generating chiral fields we
find that, $x_3 x_{17}^4=x_{40} x_3^{-1} x_{17}^2$, from which it
follows that
$$x_{40}=x_3^2 x_{17}^2,\e$$
and so we are left with three generators, $x_{17},x_{18}$ and $x_3$.
For our next relation we consider the field,
$$\theta_1=(2,1,0,0,0,0)=\theta_1-\gamma_2+\gamma_3=(1,-1,2,1,0,0).\e$$
Expressing both sides in terms of the generators we find
$x_{17}^2 x_{18}^3=x_3 x_{17}^5$, which gives us the relation,
$$x_{18}^3=x_3 x_{17}^3.\e$$
Next we take,
$$\theta_2=(0,2,1,0,0,0)=\theta_2-\gamma_3+\gamma_4=(0,1,-1,2,1,1),\e$$
which implies that $x_{18}^2 x_{17}^2=x_{18} x_{17}^{-2} x_3^2 x_{40}^2$,
which, using the expression for $x_{40}$, eq. (3.46), gives
$$x_{18}=x_3^6,\e$$
and so we are left with two generators $x_3$ and $x_{17}$. Substituting
the relation for $x_{18}$ into $x_{18}^3=x_3 x_{17}^3$ we find
the relation among the two generators,
$$x_3^{17}=x_{17}^3.\e$$
For the zero relation, we note that the first field that we miss
is at degree $139$. Writing it in terms of the generators, it gives,
$$x_{17}^2 x_3^{35}=0,\e$$
and these are the only relations in the algebra. Thus the chiral algebra
is given by,
$${\cal C}={C[x_3,x_{17}]\over (x_3^{17}-x_{17}^3, 
x_{17}^2 x_3^{35})},\e$$
and the Poincare polynomial is, since the relations are 
algebraically independent,
$$P(t^{123})={(1-t^{51})(1-t^{139})\over (1-t^3)(1-t^{17})}.\e$$
Thus, this is an ALG theory with two generators.

Let us do the example of the Dynkin diagram of $E_6$. Here the
scalar product matrix is,
$$\Gamma=\pmatrix{3&1&0&0&0&0\cr
                  1&3&1&0&0&0\cr
                  0&1&3&1&0&1\cr
                  0&0&1&3&1&0\cr
                  0&0&0&1&3&0\cr
                  0&0&1&0&0&3\cr},\e$$
and the weights are 
$$x_i=(12,10,4,10,12,14)/23\e$$ 
and the central charge is $c/3=31/23$.
We have, a priori, three generators,
$$\manyeq{x_1&=(1,-1,0,0,0,0)=(0,0,0,-1,1,0)\cr
          x_5&=(0,1,0,0,0,0)=(0,0,0,1,0,0)\cr
          x_7&=(0,0,0,0,0,1).\cr}$$
For the first relation, we consider the field,
$$\theta=(2,1,0,0,0,0)=\theta-\gamma_2+\gamma_3=(1,-1,2,1,0,1).\e$$
Expanding both sides we find, $x_1^2 x_5^3=x_7 x_5 x_1^5$ and so
the first relation is, 
$$x_5^2=x_7 x_1^3.\e$$
Taking now the field,
$$\theta_1=(0,2,1,0,0,0)=\theta_1-\gamma_3+\gamma_6=
(0,1,-1,-1,0,2),\e$$
we find that $x_1^2 x_5^2=x_7^2 x_1^{-2}$ and so the
second relation is,
$$x_7^2=x_1^4 x_5^2.\e$$
Combining the two relations, we find that $x_7^2=x_7 x_1^7$ and so
$$x_7=x_1^7,\e$$
and we are left with two generators, $x_1$ and $x_5$. The eq. (3.61)
then becomes the relation of these two generators,
$$x_1^{10}=x_5^2.\e$$
For the zero relation, we note that the first missing field is at
degree $27$. We have two fields at this degree, $x_1^{27}$ and 
$x_5 x_1^{22}$ and thus the zero relation is
$$x_1^{27}=\alpha x_5 x_1^{22},\e$$
where $\alpha$ is some unknown coefficient which depends on the
theory, but we cannot calculate as of now. 
For algebraic independence of the relations, we demand that 
$\alpha\neq \pm1,0$. To see this, note that we should not have any 
field at degree $32$. We can calculate
$$x_1^{32}=\alpha x_5 x_1^{27}=\alpha^2 x_5^2 x_1^{22}=
\alpha^2 x_1^{32}.\e$$
Thus, for $\alpha^2\neq1$, $x_1^{32}=0$. If $\alpha\neq0$ then this
equation implies that $x_5 x_1^{27}=0$ and thus we find no chiral
field at degree $32$, as we should.
Thus, the chiral algebra
is,
$${\cal C}\approx {C[x_1,x_5]\over (x_1^{10}-x_5^2,x_1^{27}-\alpha 
x_1^{22} x_5)},\e$$
and since the two relations are algebraically independent, the 
Poincare polynomial is,
$$P(t^{23})={(1-t^{10})(1-t^{27})\over (1-t)(1-t^5)}.\e$$
Again, this theory is an ALG with two generators.
 
Now, let us demonstrate that the vast majority of these models are
not rational. To do this consider the model with the $\Gamma$ matrix,
$$\Gamma=\pmatrix{3&1&0&0&0&0\cr
                 1&3&1&1&0&0\cr
                 0&1&3&1&1&0\cr
                 0&1&1&3&1&0\cr
                 0&0&1&1&3&1\cr
                 0&0&0&0&1&3\cr},\e$$
whose $x_i$ are given by 
$$x_i=(6,2,4,4,2,6)/10.\e$$
The central charge is given by $c=18/5$, so the model is, of course,
not minimal.
In this model, consider the field 
$$a=(0,-1,1,0,0,0).$$
This field mixes only with the field
$$b=a+\gamma_2-\gamma_3=(1,1,-1,0,-1,0).\e$$
The matrix elements for this mixing are given by
$$M_{11}={1\over4}\sum x_i \epsilon_i^2={1\over4}(x_2+x_3)=
{3\over20}.\e$$  
and
$$M_{22}={1\over4}(x_1+x_2+x_3+x_5)={7\over20}.\e$$
and
$$M_{12}=M_{21}={1\over2}\sqrt{x_2 x_3}={\sqrt{2}\over10}.\e$$
The dimensions are given by the eigenvalues of the matrix
$$M=\pmatrix{M_{11}&M_{12}\cr M_{21}&M_{22}\cr},\e$$
which can be seen to be
$$\Delta={5\pm 2\sqrt{3}\over20},\e$$
which are clearly not rational. We conclude that this model,
like most other IBM, is not rational. Of course, the chiral fields
will have rational dimensions since the $U(1)$ charges are rational,
and the dimensions of these fields is half the $U(1)$ charges.
 
Let us calculate the Chiral algebra of this non--rational model.
We have three generators
$$\manyeq{x_2&=(1,-1,0,0,0,0)=(0,0,0,0,-1,1)\cr
          x_1&=(0,1,0,0,0,0)=(0,0,0,0,1,0)\cr
          z_2&=(0,0,1,0,0,0)=(0,0,0,1,0,0)\cr}$$
The field $x_2=x_2-\gamma_1+\gamma_2=(-1,1,1,1,0,0)$ and so
$x_2^2=(0,0,1,1,0,0)=z_2^2$ and so the first relation is
$$x_2^2=z_2^2.\e$$
For the second relation, we take,
$$\theta=(2,1,0,0,0,0)=\theta-\gamma_2+\gamma_3=(1,-1,2,0,1,0),\e$$
from which it follows that $x_2^2 x_1^3=x_1 x_2 z_2^2$, or
$$x_2=x_1^2,\e$$
and we are left with two generators, $x_1$ and $z_2$. 
The generators obey the relation,
$$x_1^4=z_2^2,\e$$
and it can be seen to be the only relation.
For the zero relation, we note that the first missing field
is at degree $Q=11$. Since we have two generators at this level,
the zero relation becomes,
$$x_1^{11}=\beta z_2 x_1^{9},\e$$
where $\beta$ is some number not equal $\pm1$ or $0$
(for the algebraic independence, with a similar calculation as for the
$E_6$ example). 
Since the theory is not rational,
$\beta$ is, presumably, not rational. Thus, the chiral algebra
is not a ring, in this case, as we expect for most irrational theories,
but an algebra over the complex numbers.
Of course, calculating $\beta$, directly, is an extremely difficult
task. The chiral algebra thus assumes the form,
$${\cal C}\approx {C[x_1,z_2]\over (z_2^2-x_1^4,x_1^{11}
-\beta z_2 x_1^9)},\e$$
and the Poincare polynomial is, since the relations are algebraically
independent,
$$P(t^{10})={(1-t^4)(1-t^{11})\over (1-t)(1-t^2)}.\e$$
Again this is an ALG with two generators. The dimensions in all IBM
are of course algebraic numbers, since they are the eigenvalues
of an algebraic matrix. We call such irrational theories algebraic.
All the superconformal IBM's are thus algebraic theories. Presumably,
this also implies that $\beta$ is an algebraic number. I.e., we
conjecture that the constants in the chiral algebra for a rational
algebra are rational (in the appropriate basis) 
and for an algebraic theory they are algebraic.

Let us calculate a similar example. This is the graph with the scalar
product matrix,
$$\Gamma=\pmatrix{3&1&0&0&0&0&0\cr
                  1&3&1&1&1&0&0\cr
                  0&1&3&1&1&1&0\cr
                  0&1&1&3&1&1&0\cr
                  0&1&1&1&3&1&0\cr
                  0&0&1&1&1&3&1\cr
                  0&0&0&0&0&1&3\cr},\e$$
whose weight vector is 
$$x_i=(14,2,8,8,8,2,14)/22.\e$$
The central
charge is $c/3=28/22$. The generators of the chiral algebra are,
$$\manyeq{x_6&=(1,-1,0,0,0,0,0)=(0,0,0,0,0,-1,1)\cr
          x_1&=(0,1,0,0,0,0,0)=(0,0,0,0,0,1,0)\cr
          z_4&=(0,0,1,0,0,0,0)=(0,0,0,1,0,0,0)=(0,0,0,0,1,0,0)\cr}$$
For $x_6$ we have, 
$$x_6=x_6+\gamma_2-\gamma_1=(-1,1,1,1,1,0,0),\e$$
which implies that
$$x_6^2=z_4^3,\e$$
Next, we consider the field,
$$\theta_1=(2,1,0,0,0,0,0)=\theta_1-\gamma_2+\gamma_3=
(1,-1,2,0,0,1,0),\e$$
which implies that $x_6^2 x_1^3=x_6 x_1 z_4^2$, or
$$z_4^2=x_6 x_1^2.\e$$
Multiplying this equation by $z_4$ and using eq. (3.88) we find
$$x_6=x_1^2 z_4.\e$$
Substituting into eq. (3.90), we find,
$z_4^2=z_4 x_1^4$ and so
$$z_4=x_1^4,\e$$
and we are left with one generator, $x_1$, $x_6=x_1^6$. The missing
field is at the degree $Q=29$ so the chiral algebra assumes the form,
$${\cal C}\approx {C[x_1]\over (x_1^{29})},\e$$
and the Poincare polynomial is
$$P(t^{22})={1-t^{29}\over 1-t},\e$$
and the theory is an ALG with one generator. Note that although
the Poincare polynomial is similar to that of the $27$th minimal
model, this is not a minimal model, as the grading is by a different
power of $t$.

Let us do now an example with three generators. As we shall see,
with three generators we always have four relations, which are no longer
algebraically independent. Thus, the Poincare polynomial is no longer
of the simple structure eq. (3.10). However, we still call this structure
almost Landau Ginzburg (ALG), as it is common to many IBM's with
three generators.

Consider then the theory $A_7$. The scalar product matrix, $\Gamma$
is given by eq. (3.1). The weights are 
$$x_i=(26,16,20,18,20,16,26)/47.\e$$
The central charge is $c/3=71/47$. We have three generators,
$$\manyeq{x_5&=(1,-1,0,0,0,0,0)=(0,0,0,0,0,-1,1)\cr
          x_8&=(0,1,0,0,0,0,0)=(0,0,0,0,0,1,0)\cr
          x_9&=(0,0,0,1,0,0,0)\cr}$$
We know that $x_5=x_5-\gamma_1+\gamma_2=(-1,1,1,0,0,0,0)$. Consequently,
$x_5^2=(0,0,1,0,0,0,0)$. For the first relation, we consider,
$$\theta=(2,1,0,0,0,0,0)=\theta-\gamma_2+\gamma_3=(1,-1,2,1,0,0,0).\e$$
Expanding both sides we find $x_5^2 x_8^3=x_5^5 x_9$, from which
we deduce the first relation,
$$x_8^3=x_9 x_5^3.\e$$
For the second relation, we consider the field,
$$\theta_1=(0,2,1,0,0,0,0)=\theta_1-\gamma_3+\gamma_4=
(0,1,-1,2,1,0,0).\e$$
Expanding both sides in terms of the generators, we find 
$x_8^2 x_5^2=x_8 x_9^2$ so
$$x_9^2=x_8 x_5^2,\e$$
which is our second relation. These are all the field identifications.

We need to find now the zero relations. The Poincare Polynomial
for the algebra with the two field identifications is generated
by fields which are $x_5^n x_9^l x_8^m$, where $n$ is any integer,
$l=0,1$ and $m=0,1,2$. It is thus,
$$P_0={1\over 1-t^5} (1+t^9)(1+t^8+t^{16}).\e$$
The highest missing power of $t$ is at $t^{12}$. We thus need
a zero relation at $t^{71-12}=t^{59}$. The only field of degree
$59$ is $x_5^{10} x_9$. We thus conclude that the first zero 
relation is
$$x_5^{10} x_9=0.\e$$
Now, we note that in $P_0$ we have a term which is $2 t^{25}$, but
we have only one field at degree $t^{71-25}$. We thus need an additional relation
at the degree $Q=25$. This relation is, thus,
$$x_5^5-\zeta x_9 x_8^2=0,\e$$
with some coefficient $\zeta$. Closure of the algebra implies
that $\zeta=1$, as we can see by multiplying both sides with $x_8$,
and using the relations.
With these four relations the Poincare polynomial becomes
$$P(t^{47})=1+t^5+t^8+t^9+t^{10}+(t^{13}+t^{14}+\ldots + t^{58})+
t^{61}+t^{62}+t^{63}+t^{66}+t^{71},\e$$
which obeys the transpose relation, eq. (3.8), but the relations
in the algebra
are no longer algebraically independent. We conclude that the
chiral algebra is
$${\cal C}\approx {C[x_5,x_8,x_9]\over (x_8^3-x_9 x_5^3, 
x_9^2-x_8 x_5^2,x_5^{10} x_9,x_5^5-x_8^2 x_9)}.\e$$

Let us do another example with three generators, which is $A_8$.
Here the weight vector is,
$$x_i=(21,13,16,15,15,16,13,21)/38,\e$$
and the central charge is
$c/3=130/76$. We have three generators for the algebra,
$$\manyeq{x_8&=(1,-1,0,0,0,0,0,0)=(-1,1,1,0,0,0,0,0)\cr
          x_{13}&=(0,1,0,0,0,0,0,0)\cr
          x_{15}&=(0,0,0,1,0,0,0,0).\cr}$$
For the first relation, we have,
$$\theta=(2,1,0,0,0,0,0,0)=\theta-\gamma_2+\gamma_3=
(1,-1,2,1,0,0,0,0).\e$$
Expressing this relation in terms of the generators gives 
$x_8^2 x_{13}^3=x_8^5 x_{15}$ which implies that
$$x_{13}^3=x_{15} x_8^3,\e$$
which is the first relation at degree $39$.

Next, we take
$$\theta_1=(0,2,1,0,0,0,0)=\theta_1-\gamma_3+\gamma_4=
(0,1,-1,2,1,0,0,0),\e$$
which implies that $x_{13}^2 x_8^2=x_{13} x_8^{-2} x_{15}^3$.
This gives us our second field identification,
$$x_{15}^3=x_{13} x_{8}^4,\e$$
at degree $45$.

To get the zero relations we write down the Poincare polynomial of the
two field identifications. This is generated by fields of the
type $x_8^n x_{13}^l x_{15}^m$, where $n$ is any integer and 
$l$ and $m$ equal $0,1$ or $2$. Thus, the Poincare polynomial is
$$P_0(t^{76})={1\over 1-t^8} (1+t^{13}+t^{26})(1+t^{15}+t^{30}).\e$$
The maximal missing power of $t$ in $P_0$ is $t^{35}$. We thus
need a relation at the degree $Q=Q_{\rm max}-35=95$. This gives us
the first zero relation,
$$x_{15} x_8^{10}=0.\e$$
We have two fields in $P_0$ at degree $56$. One of them 
needs to be eliminated
since we have only one field at degree $130-56=74$. We thus find the
second zero relation,
$$x_8^7=x_{15}^2 x_{13}^2,\e$$
where the coefficient was determined by closure of the algebra 
(multiplying by $x_{15}$ or $x_{13}$ and using the relations).
There are no other relations. We conclude that the chiral algebra is
$${\cal C}\approx {C[x_8,x_{13},x_{18}]\over 
(x_{13}^3-x_8^3 x_{15},x_{15}^3-x_8^4 x_{13},x_{15} x_8^{10},
x_8^7-x_{15}^2 x_{13}^2)},\e$$
and the Poincare polynomial is
$$P(t^{76})=P_1(t)+t^{130} P_1(1/t)+\sum_{r=36}^{94} t^r,\e$$
where
$$P_1(t)=1+t^8+t^{13}+t^{15}+t^{16}+t^{21}+t^{23}+t^{24}+t^{26}+t^{28}+
t^{29}+t^{30}+t^{31}+t^{32}+t^{34}.\e$$
Again, the relations are not algebraically independent, but we have
four relations with only three generators.
\par
\mysec{Conclusions}
We discussed here the chiral algebras of $N=2$ superconformal 
interacting boson models. We have seen that a common structure arises
which we called almost Landau Ginzburg. We hope that these results
will lead to further understanding of the geometrical interpretation
of these models.

The examples discussed here were chosen rather sporadically. The calculus
we developed works equally well on most other superconformal 
interacting bosons,
which, indeed, is an enormous wealth of theories.

One can describe a string compactification on tensor products of 
such IBM's, along the lines of ref. \r\DG. For four dimensional string
theories, the only requirement is that the total central charge should 
be $9$. Although, most IBM's are not rational, this is not a problem,
and the construction of ref. \r\DG, works equally well for those. 
The chiral
algebras discussed here would then give the number of generations and
anti--generations in such string theories, along with the Yukawa 
couplings. Thus, such models can be of phenomenological interest.   
\refout 
\bye

%% file: mydef.tex
\def\e{\adveq\eqno{\rm (\chapterlabel\the\equanumber)}}
\def\mysec#1{\equanumber=0\chapter{#1}}
\def\adveq{\global\advance\equanumber by 1}
\def\myeq{{\rm \chapterlabel\the\equanumber}}
\def\rarrow{\rightarrow}

\def\manyeq#1{\eqalign{#1}\e}

\def\dim{\mathop{\rm dim}\nolimits}

\def\semidirect{\mathrel{\raise0.04cm\hbox{${\scriptscriptstyle |\!}$
\hskip-0.175cm}\times}}


\def\ref#1{$^{[#1]}$}

\def\pr#1{#1^\prime}
 
\def\r#1{$[\rm#1]$}

\def\e{\adveq\eqno{\rm (\chapterlabel\the\equanumber)}}
\def\mysec#1{\equanumber=0\chapter{#1}}
\def\adveq{\global\advance\equanumber by 1}
\def\myeq{{\rm \chapterlabel\the\equanumber}}
\def\rarrow{\rightarrow}

\def\manyeq#1{\eqalign{#1}\e}

\def\dim{\mathop{\rm dim}\nolimits}

\def\semidirect{\mathrel{\raise0.04cm\hbox{${\scriptscriptstyle |\!}$
\hskip-0.175cm}\times}}


\def\ref#1{$^{[#1]}$}

\def\pr#1{#1^\prime}
 
\def\r#1{$[\rm#1]$}

\def\half{{1\over2}}